\documentclass[prl,twocolumn,showpacs,groupedaddress]{revtex4}

\usepackage{amsfonts}
\usepackage{stmaryrd}
\usepackage{amssymb}
\usepackage{graphics,epsfig}
\usepackage{bm}
\usepackage{subeqn}

\begin{document}

\title{Coherence Frame, Entanglement Conservation, and Einselection}

\author{Dong-Sheng Wang}
\email{wdsn1987@gmail.com}

\date{26 September 2011}

\begin{abstract}
In this paper, the theory of coherence frame is developed. Two kinds
of coherence frame are classified. Under coherence frame, the
entanglement is conserved in the entanglement swapping process,
without entanglement sudden death and birth. The einselection method
for the preferred basis problem in the entangle process is shown as
incomplete.
\end{abstract}


\maketitle

Entanglement, the foundation of quantum dynamics and quantum
information processing (QIP), is realized as a special kind of
quantum correlation \cite{Horodecki}. Many seminal arguments and
concepts have been induced by entanglement in various contexts,
e.g., the EPR paradox with hidden variable theory \cite{EPR,bell},
the collapse for measurement \cite{neumann}, the superselection rule
(SSR) \cite{www} etc. Recently, the method of coherence frame (CF)
(or quantum reference frame) \cite{Rudolph} is widely concerned in
QIP, e.g., clock synchronization, phase reference etc. It proves
that the method, einselection \cite{zurek1,zurek2}, to resolve the
preferred basis problem, and the method, the Aharonov-Susskind
experiment \cite{as}, to challenge the SSR are based on the method
of CF. In this Letter, we focus on the theory of CF relating to
entanglement transfer and conservation.

The CF problem rises when the representation can not be easily
decided, which is traditionally seldom analyzed \cite{Tannoudji}.
For single-body system, e.g., spin $S_{1/2}$, only one
representation can exist in reality at one time. If under the
representation of $\hat{S}_x$, the measurement on $\hat{S}_y$ or
$\hat{S}_z$ is meaningless. Yet, for double $S_{1/2}$ system, any
component of the first spin $\hat{S}_1$ commutates with that of the
second spin $\hat{S}_2$ \cite{Tannoudji}. Thus, $\{\hat{S}_1,
\hat{S}_2\}$ forms the complete set of commuting observable (CSCO).
In the entangled state $|\Phi\rangle=(|\uparrow_z \downarrow_x
\rangle+|\downarrow_z \uparrow_x \rangle)/\sqrt{2}$, spin
$\hat{S}_1$ ($\hat{S}_2$) is under the $\hat{S}_z$ ($\hat{S}_x$)
representation, the $z$-component of $\hat{S}_1$ and $x$-component
of $\hat{S}_2$ can be simultaneously defined. Another way to extend
representation is to consider the time evolution, leading to the
method of decohered history, i.e., framework \cite{griffiths}. Under
one framework, the representation may differ in various sections of
the history, e.g., $\hat{S}_x$ and $\hat{S}_z$ can be measured
successively for single spin.

The theory of CF mainly involves the many-body and macroscopic
superposition and dynamics, for which, the SSR was developed earlier
\cite{www}. Lacking one CF, leading to decoherence, is equivalent to
SSR. Also, the method of SSR is shown similar with density matrix
for mixed state \cite{www}. Note that there are two cases for SSR.
Case (1): SSR for isotropic particles, e.g., electrons, which is the
ensemble. Case (2): SSR for particles not isotropic, e.g., electrons
and protons, which is the mixture. The Hilbert space is the direct
sum of that of each kind of particle, and the density matrix is
block-diagonal. As shown in Ref. \cite{as}, it is possible to
introduce coherence to the mixture. Thus, the difference of the two
cases is merely apparent, and we need only refer to the first case
for simplicity. Below, we discuss the relation of CF with SSR and
density matrix in detail.

The density matrix is $\rho=\sum_ip_i|\psi_i\rangle\langle\psi_i|$,
$\sum_ip_i=1$, $|\psi_i\rangle$ is the state vector for the
subsystem. In the spirit of SSR, the state vector of the whole
ensemble can be \cite{Wang1}
\begin{equation}
|\Psi\rangle=\sum_i\gamma_i|\psi_i\rangle,
\end{equation}
where the parameters $\gamma_i$ are complex, with $|\gamma_i|^2=p_i$.

The density matrix can be re-written as
\begin{subeqnarray}
\label{eq:en}
\label{eq:ena}
\rho\equiv|\Psi\rangle\langle\Psi|&=&\sum_i|\gamma_i|^2|\psi_i\rangle\langle\psi_i|\\
\label{eq:enb}
&\neq&\sum_{i,j}\gamma_i\gamma_j|\psi_i\rangle\langle\psi_j|,
\end{subeqnarray}
with $|\langle\psi_i|\psi_j\rangle|\geq 0$, the nonorthogonality
condition.

The illegality of the form in Eq.~(\ref{eq:enb}) is ensured by the
SSR; if not, the coherence in the ensemble will be considered
redundantly. Physically, the form in Eq.~(\ref{eq:ena}) is a kind of
coarse-graining, since each sector $|\psi_i\rangle$ can be further
decomposed as the superposition of orthogonal eignstates
\cite{Wang1}. This indicates that the coherence due to
nonorthogonality is localized and {\em globally inaccessible}. To
surpass SSR, the CF and special interactions are needed to entangle
the sectors together, as a result, the localized coherence will
transfer globally and realize the coherence delocalization
\cite{zurek3}. In the entangled state, parties can be the CF of each
other, respectively. Tracing out one party, equivalent to SSR
physically, leads to the apparent decoherence and classical state.
Note that when SSR applies, the state is not necessarily classical.

To develop the complete theory of CF, primarily, we propose that CF
should be clarified into several categories since there exist
different quantum coherence and entangle processes. Briefly, we
separate two kinds of CF:

(I). {\em Primary coherence frame}: quantum field, vacuum,
environment (like noise, phonon, etc).

The role of the primary CF in some cases may be trivial then can be
ignored. This type plays the similar role as space-time in classical
mechanics.

(II). {\em Measurement-type coherence frame}: various interacting
systems under specific conditions.

This type includes all kinds of CF beyond the primary type, in
principle. In practice, this type mainly refers to the parties in
the entangled or nonlocal states.

For clarity, we take the well known double-slit interference
experiment for example. The electron ensemble shows fringes under
the structure of double-slit. The primary CF of electron is the
electromagnetic field, which, together with electron, form the
entangled state
\begin{subequations}
\begin{equation}
|\psi\rangle=\alpha_1|l\rangle|f_l\rangle+\alpha_2|r\rangle|f_r\rangle,
\end{equation}
where $|l(r)\rangle$ ($|f_{l(r)}\rangle$) stand for the eignstates
of electron (field) at the left and right slits,
$\alpha_1^2+\alpha_2^2=1$. Physically, the electromagnetic field is
the continuous variable system, i.e., $|f_l\rangle \approx
|f_r\rangle \equiv |f\rangle$, thus, the state reduces to
\begin{equation}
|\psi\rangle=(\alpha_1 | l \rangle+\alpha_2 | r \rangle)|f\rangle,
\end{equation}
\end{subequations}
it is clear that the state of electron is superposed leading to the
interference. The CF in this case is {\em trivial}, that it does not
delocalize the coherence of electron to the field.

If we introduce measurement, which will interact with the electron
thus become one of the parties of the global state. The state under
measurement-type CF is
\begin{equation}
|\phi\rangle=(\beta_1|l\rangle|M_l\rangle+\beta_2|r\rangle|M_r\rangle)|f\rangle,
\end{equation}
where $|M_{l(r)}\rangle$ are the states of the measurement,
$\beta_1^2+\beta_2^2=1$. The interfere process depends on the type
of measurement, e.g., weak measurement \cite{Braginsky}, under which
both the wave and particle properties can be observed.

The phenomenon for macroscopic objects, e.g., the ball, however, is
different \cite{Feynman}. Note that the CF for this case should be
the gravitational field relating to the classical dynamics. The
entangled state reads
\begin{subequations}
\begin{equation}
|\chi\rangle=\gamma_1|l\rangle|c_l\rangle+\gamma_2|r\rangle|c_r\rangle,
\end{equation}
where $|c_{l(r)}\rangle$ stand for the eignstates of field at the
left and right slits, $\gamma_1^2+\gamma_2^2=1$. Since states
$|c_l\rangle$ and $|c_r\rangle$ can be distinguished via Newtonian
mechanics, i.e., $\langle c_l | c_r\rangle=0$, the state of the ball
should be density matrix
\begin{equation}
\rho_b=\textrm{tr}_c(|\chi\rangle\langle\chi|)=\gamma_1^2|l\rangle\langle
l|+\gamma_2^2|r\rangle\langle r|,
\end{equation}
\end{subequations}
which brings the classical results and the absence of interference.
In addition, this does not mean the movement in gravitational field
is classical, in contrast, the corresponding effects of quantum
coherence can be significant on the larger scale in Quantum
Cosmology \cite{Halliwell}.

{\em Entanglement transfer under CF.} Next, we explore the basic
entanglement transfer (re-distribution) process, a kind of coherence
delocalization, under CF. It is reasonable to assume that for one
system there exists at least one CF, which leads to the bi-party
entanglement. This is similar with the so-called ``pure universe''
model for other studies \cite{Popescu,Wang2}. Under the Schmidt
representation, the entangled state is
\begin{subequations}
\begin{equation}
|\Psi\rangle=\sum^n_i\lambda_i|s_i^A\rangle|e_i^A\rangle,
\end{equation}
where $|s_i^A\rangle$ ($|e_i^A\rangle$) is the basis for the system
(referred as) $A$ (CF). If $|\langle e_i|e_j \rangle|\approx 1$,
i.e., the disturbance and reference effect of the CF to the system
is trivial, thus, the CF can be treated classically, which is a kind
of classical limit satisfying the correspondence principle
\cite{Bohr}.

Suppose, there exists another system (referred as $B$) which will be
entangled together with the system $A$. The state of system $B$ with
its CF is expressed as
\begin{equation}
|\Phi\rangle=\sum^m_j\gamma_j|s^B_j\rangle|e^B_j\rangle,
\end{equation}
\label{eq:cf}
\end{subequations}
where $|s^B_j\rangle$ ($|e^B_j\rangle$) is the basis for the system
$B$ (CF). The general bi-party entangle process under CF takes as
\begin{equation}
|\Psi\rangle|\Phi\rangle \rightarrow
\sum^w_k\alpha_k|s^A_k\rangle|s^B_k\rangle\sum^w_k\beta_k|e^A_k\rangle|e^B_k\rangle,
\label{eq:basiccf}
\end{equation}
with $w\leq \textrm{min}(n,m)$, the coefficients $\lambda_i$,
$\gamma_j$, $\alpha_k$, $\beta_k$ guarantee the normalization rule.
This process is similar with the entanglement swapping process
firstly realized by Bell-type states \cite{Bennett}. The von Neumann
pre-measurement for the collapse model \cite{neumann} ignores the
roles of environment and quantum openness \cite{zeh,Wang2}, i.e.,
ignoring the CF, which can be deduced by taking the classical limit
of the swapping process.

Further, we study the entanglement transfer quantitatively. One
central problem is that too many entanglement measures exist
especially for mixed state \cite{plenio}. The main physical reason
is there does exist the mismatch between entanglement and density
matrix. Entanglement, as the many-body property, is defined
referring to the number of parties, while density matrix is defined
referring to state instead of parties. As a result, the expression
{\em entanglement in density matrix} is not complete. Below, we
present our method of entanglement measure. Generally, density
matrix can be two types: ensemble of single-body system and ensemble
of many-party system (with or without noise), only the later type
contains entanglement. For the bi-party system, the
ensemble-entangled qudit \cite{Wang1} is written as
\begin{eqnarray}
\rho_{eE}&=&\sum_{\xi} p_{\xi} |\psi_{\xi}\rangle \langle
\psi_{\xi}| \\ \nonumber &=& \sum_{\xi} p_{\xi} \sum_{i,j}
\lambda^{\xi}_i \lambda^{\xi}_j |A_i\rangle \langle A_j
|\otimes|B_i\rangle \langle B_j |,
\end{eqnarray}
with $\sum_i (\lambda^{\xi}_i)^2=1$, $\sum_{\xi} p_{\xi}=1$,
$|A_i\rangle$ ($|B_i\rangle$) is the local eigenstate of subsystem
$A$ ($B$), $\lambda^{\xi}_i$ is Schmidt coefficient.

When $\xi=1$, the state reduces to the pure entangled qudit. The
degree of entanglement \cite{Wang1} for the pure entangled qudit
state is defined as
\begin{equation}
E \equiv \sum_{i<j}^n |\lambda_i||\lambda_j|, \label{eq:entan}
\end{equation}
which is the entanglement monotone \cite{plenio}. This entanglement
measure characterizes the {\em distributed coherence}, different
from information (entropy).

For the general mixed state $\rho$, the method of decomposition is
employed to extract its entanglement
\begin{equation}
\rho=\pi \rho_{eE} + (1-\pi) \rho', \label{eq:decom}
\end{equation}
with $\pi\in [0,1]$. The state $\rho$ can be viewed as the result of
the initial state $\rho_{eE}$ disturbed by $\rho'$, thus, the
entanglement in $\rho$ is defined as $E(\rho)=\pi E(\rho_{eE})$. We
name state $\rho_{eE}$ as the {\em natural point} of state $\rho$
for convenience.

Physically, there exist problems of the entanglement of formation
$E_f$ \cite{Wootters}, which relies on the decomposition to
entangled qudit. However, from Eq. (\ref{eq:decom}), there are
situations noise or product states ($\rho'$) are added, so that
$E_f$ and the related concurrence do not directly quantify
entanglement. The detailed properties of this measure have partly
been discussed in Ref. \cite{Wang1}.

For example, Werner state $|\rho_w\rangle=\frac{1-z}{4}\textrm{I}+z
|\Psi^-\rangle\langle \Psi^-|$ is naturally the mixture of noise and
singlet, the entanglement is just $z/2$. For the X-type state with
$\rho_{14}=\rho^*_{41}=w$, $\rho_{23}=\rho^*_{23}=z$, the
entanglement equals $|w|+|z|$.

For convenience, we model system $A$ and $B$ also their CF
(environment) $\alpha$ and $\beta$ as qubit. Based on Eq.
(\ref{eq:cf}), the initial state for the global four-party system is
set as
\begin{eqnarray}
|\Upsilon\rangle &=&
|\psi_{A\alpha}\rangle\otimes|\psi_{B\beta}\rangle
\\ \nonumber
&=& ( a_1 |s^A_1e^A_1\rangle + a_2 |s^A_2e^A_2\rangle) \otimes ( b_1
|s^B_1e^B_1\rangle + b_2 |s^B_2e^B_2\rangle ),
\end{eqnarray}
after swapping, which can also be expressed as
\begin{equation}
|\Upsilon\rangle=|\psi^+_{AB}\rangle|\psi^+_{\alpha\beta}\rangle+|\psi^-_{AB}\rangle|\psi^-_{\alpha\beta}\rangle
+|\phi^+_{AB}\rangle|\phi^+_{\alpha\beta}\rangle+|\phi^-_{AB}\rangle|\phi^-_{\alpha\beta}\rangle,
\end{equation}
where
\begin{eqnarray}
|\psi^{\pm}_{AB}\rangle &=& s_1 |s^A_1s^B_1\rangle\pm s_2
|s^A_2s^B_2\rangle, \\ \nonumber |\phi^{\pm}_{AB}\rangle &=& t_1
|s^A_1s^B_2\rangle\pm t_2 |s^A_2s^B_1\rangle, \\ \nonumber
|\psi^{\pm}_{\alpha\beta}\rangle &=& x_1 |e^A_1e^B_1\rangle\pm x_2
|e^A_2e^B_2\rangle, \\ \nonumber |\phi^{\pm}_{\alpha\beta}\rangle
&=& y_1 |e^A_1e^B_2\rangle\pm y_2 |e^A_2e^B_1\rangle,
\end{eqnarray}
the coefficients $a_i$, $b_i$, $s_i$, $t_i$, $x_i$, and $y_i$
($i=1,2$) satisfy the normalization rule, also, $a_1b_1=2s_1x_1$,
$a_2b_2=2s_2x_2$, $a_1b_2=2t_1y_1$, $a_2b_1=2t_2y_2$. Under
symmetrical condition, it reduces to the swapping for Bell's state
\cite{Bennett}.

From the definition in Eq. (\ref{eq:entan}), the entanglement of
state, e.g., $|\psi_{A\alpha}\rangle$ is $|a_1a_2|$. It is easy to
find that the entanglement during the swapping satisfies
\begin{eqnarray}
\label{eq:swap}
&E(|\psi_{A\alpha}\rangle)E(|\psi_{B\beta}\rangle)=\;\;\;\;\;\;\;\; \;\;\;\;\;\;\;\;\;\;\;\;\;\;\;\;\;\;\;\;\;\;\;\;\;\;\;\;\;\;\;\;\;\;\;\;\;\;\;\;&\\
\nonumber
&\;\;E(|\psi^+_{AB}\rangle)E(|\psi^+_{\alpha\beta}\rangle)+E(|\psi^-_{AB}\rangle)E(|\psi^-_{\alpha\beta}\rangle)
&\\ \nonumber
&+E(|\phi^+_{AB}\rangle)E(|\phi^+_{\alpha\beta}\rangle)+E(|\phi^-_{AB}\rangle)E(|\phi^-_{\alpha\beta}\rangle),&
\end{eqnarray}
calculated as $|a_1a_2b_1b_2|=4|t_1t_2y_1y_2|=4|s_1s_2x_1x_2|$,
which can be verified by the coefficients relation above. If we view
the entanglement of the four-party state $|\Upsilon\rangle$ as the
product of the entanglement of the bi-party state
$|\psi_{A\alpha}\rangle$ and $|\psi_{B\beta}\rangle$, the relation
in Eq. (\ref{eq:swap}) stands for a kind of {\em conservation of
entanglement} during the swapping. If the initial state of the
global state is other types, it is easy to check that the
conservation still exists. For the general mixed state, based on the
decomposition in Eq. (\ref{eq:decom}), the entanglement is also
conserved.

As the example, we study one actual system from Ref. \cite{lopez}
relating to the entanglement sudden death and birth (ESDB)
\cite{yu}. The model contains the entangled cavity photons affected
by dissipation (i.e., CF). The initial state is set as
$|\Phi_0\rangle=(\alpha|0\rangle_{c_1}|0\rangle_{c_2}+\beta|1\rangle_{c_1}|1\rangle_{c_2})|{\bf
\bar{0}}\rangle_{r_1}|{\bf \bar{0}}\rangle_{r_2}$, the entanglement,
amount to $|\alpha\beta|$, is distributed between the two cavity
photons. The reduced density matrix of the two-cavity is
$\rho_{c_1c_2}$ \cite{lopez}, from Eq. (\ref{eq:decom}), it is
direct to get its natural point $\sigma$, written by elements,
$\sigma_{11}=\alpha^2/\pi$, $\sigma_{44}=\beta^2\xi^4/\pi$,
$\sigma_{14}=\sigma_{41}=\alpha\beta\xi^2/\pi$, others zero. The
parameter $\pi=\alpha^2+\beta^2\xi^4$,
$\xi=\sqrt{1-\chi^2}=\exp(-\kappa t/2)$, $\kappa$ is the decay
constant. Thus, the entanglement of state $\rho_{c_1c_2}$ is
$E(\rho_{c_1c_2})=|\alpha\beta|\xi^2$. Following the similar method,
the entanglement for the state of the two-reservoir is
$E(\rho_{r_1r_2})=|\alpha\beta|\chi^2$. Thus, the conservation reads
\begin{equation}
E(|\Phi_0\rangle)=E(\rho_{c_1c_2})+E(\rho_{r_1r_2})=|\alpha\beta|.
\end{equation}

As time evolves, $E(\rho_{c_1c_2})$ of the two-cavity decreases
exponentially, while $E(\rho_{r_1r_2})$ of the two-reservoir
increases exponentially. There is no ESDB, i.e., the localized and
distributed coherence translate mutually.

{\em Preferred basis problem.} From the method of CF, we discuss the
preferred basis problem (PBP), which has been studied via the
einselection approach \cite{zurek1,zurek2}. We will show, yet, the
method of einselection is incomplete.

This method is described via the Stern-Gerlach experiment, as shown
in the Fig. $1$ of Ref. \cite{zurek1}. The system is represented by
the spin states (up and down) along some directions. One atom is put
near one channel to serve as the apparatus to interact with the
spin, causing entanglement. In this measurement, the PBP means that
there is no physical difference between states
\begin{eqnarray}
\label{eq:pbp} |\psi_1\rangle
&=&(|d\rangle|U\rangle-i|u\rangle|D\rangle)/\sqrt{2},
\\ \nonumber|\psi_2\rangle
&=&(|S^+\rangle|A^+\rangle-|S^-\rangle|A^-\rangle)/\sqrt{2},
\end{eqnarray}
where $|S^{\pm}\rangle=(|u\rangle \pm i|d\rangle)/\sqrt{2}$,
$|A^{\pm}\rangle=(|U\rangle \pm |D\rangle)/\sqrt{2}$, $|u\rangle$
(up), $|d\rangle$ (down), $|S^{\pm}\rangle$ are the states of the
spin $S$, $|U\rangle$, $|D\rangle$, $|A^{\pm}\rangle$ are the states
of the atom $A$.

The main ideas of einselection are as follows. Suppose the
pre-measurement between $S$ and $A$, and one special interaction
between $A$ and $E$, the global state vector is
\begin{equation}
|\phi\rangle=(-ic|uD\mu\rangle-s|uD\delta\rangle+c|dU\mu\rangle+is|dU\delta\rangle)/\sqrt{2},
\end{equation}
with $|\mu\rangle$, $|\delta\rangle$ the states of $E$, $c \equiv
\cos A(t)$, $s \equiv \sin A(t)$, $A(t)$ depends on the coupling
periodically.

Introduce $|E^{\pm}\rangle=(|\mu\rangle \pm
i|\delta\rangle)/\sqrt{2}$, for the special case $A(t)=\pi/4$, the
state reduces to the GHZ-type
\begin{equation}
|\phi\rangle=(|dUE^-\rangle-i|uDE^+\rangle)/\sqrt{2}.
\end{equation}

Operators $|U\rangle\langle U|$, $|D\rangle\langle D|$ can project
out the two branches of $|\phi\rangle$, yet, not the case for
$|A^{\pm}\rangle\langle A^{\pm}|$. Thus, states $|U\rangle$ and
$|D\rangle$ instead of $|A^{\pm}\rangle$ are the {\em pointer
states}, i.e., the environment ($E$) of $A$ serves to singlet out
the pointer state of $A$ to entangle with $S$.

The einselection method is flawed, however. One reason is that the
demonstration of pointer state depends on the special case
$A(t)=\pi/4$; in contrast, pointer state, robust to the noise,
should not depend on time, i.e, the other basis, $|A^{\pm}\rangle$,
of the apparatus cannot act due to the action of $E$. Yet, it is
obvious to check that when $A(t)=0$ or $\frac{\pi}{2}$, $A$ can be
written in $|A^{\pm}\rangle$ basis, which is in conflict with the
spirit of einselection.

Another reason is the collapse in the bi-party model for measurement
\cite{neumann} is not physical, i.e., the branches in the entangled
state only manifest the potential results of $S$ corresponding to
different states of $A$. The apparatus is a kind of {\em inner
observer}. By introducing $E$, similar with the physical collapse
theory \cite{PCT}, einselection improperly changes entanglement from
bi-party to tri-party type.

Indeed, the PBP can be resolved by the CF method. We re-express the
PBP into two aspects:

(I). {\em Whether the entangled state between one system $S$ and
another system $A$ can be written in infinite ways as}
\begin{subequations}
\begin{equation}
|\psi\rangle=\sum_i a_i |S_i\rangle |A_i\rangle=\sum_j b_j
|S'_j\rangle |A'_j\rangle=\cdots,
\end{equation}

(II). {\em Whether the entangled state can be written as}
\begin{equation}
|\psi\rangle=\sum_i \alpha_i |S_i\rangle |A'_i\rangle=\sum_j \beta_j
|S'_j\rangle |A''_j\rangle=\cdots,
\end{equation}
where $|S_i\rangle$, $|A_i\rangle$, $\cdots$ satisfy the
superposition principle, and $a_i$, $b_j$, $\alpha_i$, $\beta_j$
satisfy the normalization rule.
\end{subequations}

PBP(I) concerns the problem of the definite form of the actual
basis, which is similar with the traditional version of PBP; while
PBP(II) concerns the problem of the exact structure of the entangled
state, i.e., how the basis of $S$ and $A$ correlate with each other
one-to-one.

Different forms in PBP(I) have special CF, relating to different
physical conditions. The states $|\psi_1\rangle$ and
$|\psi_2\rangle$ in Eq. (\ref{eq:pbp}) correspond to different
states of CF (see Eq. (\ref{eq:basiccf})). The configuration space,
whose effect to the coherence is trivial though, plays the role of
primary CF for the spin state (up and down), and decides the
representation of the system. As a result, PBP(I) is only one {\em
pseudo} problem. Note that it indicates when the CF is trivial, the
entangled state shows symmetry under rotation.

For PBP(II), the reason for the non-equivalence of various forms is
that the parties in the entangled state play the CF of each other,
respectively. The rotation of the basis of one party will change the
structure of the state, thus change the reference relation between
$S$ and $A$. The detailed form of the entangled state also depends
on the interaction within and the practical arrangement, e.g., the
generations of the two type Bell's basis are different.

Last, we note that our method of CF is consistent with the recently
developed method of {\em relational} Hilbert space \cite{Rudolph},
according to which, there exist external (also classical) and
internal CF. Under the internal CF, to extract the coherence of the
system only, the Hilbert space is mapped onto
$\mathcal{H}_{gl}\otimes\mathcal{H}_{rel}$, where $\mathcal{H}_{gl}$
plays the role of the classical CF, and $\mathcal{H}_{rel}$, the
relational Hilbert space, is the space for the system with a slight
difference from $\mathcal{H}_S$, which can be ignored in the
classical limit. That is to say, the CF should always be treated
internally, while in practice, some CF is trivial then can be
ignored.

In conclusion, we mainly developed the theory of coherence frame in
this paper. We showed that the entanglement is conserved in the
swapping process, and the entanglement sudden death and birth does
not exist. We also demonstrated that the preferred basis problem can
be resolved more naturally by the method of coherence frame than the
einselection method.

\end{document}